\documentclass[twocolumn,showpacs,preprintnumbers,amsmath,amssymb]{revtex4}
\usepackage{epsfig,amssymb}
\begin{document}

%\draft

\title{Partially Locked States in Coupled Oscillators due to Inhomogeneous Coupling} 

\author{Tae-Wook Ko}
\email{taewook@pitt.edu}
\author{G. Bard Ermentrout}
\email{bard@math.pitt.edu}
\affiliation{Department of Mathematics, University of Pittsburgh, Pennsylvania 15260, USA}
\date{\today}
\begin{abstract}
We investigate coupled identical phase oscillators with scale-free distribution of coupling strength. It is shown that partially locked states can occur due to the inhomogeneity in coupling and some properties of the coupling function. Various quantities of the partially locked states are computed through a self-consistency argument and the values show good agreement with simulation results. 
\pacs{05.45.Xt, 89.75.-k, 87.19.La}
% coupled oscillators, complex systems, neuroscience
%\\
\end{abstract}

\maketitle
\section{Introduction}
Synchronization of coupled oscillators is important and has been
widely studied in variety of systems from physics, chemistry and biology
\cite{pikovsky_book,sync,kura,strogatz_physica,kura_review,ermentrout2001}. 
Partially locked states are states of imperfect synchronization with locked subpopulation and drifting subpopulation, and have been observed in coupled oscillators with distributed intrinsic frequencies \cite{kura,strogatz_physica,kura_review}. 
The formation of locked subpopulation with the increase of the coupling strength is one of the explanations for the transition between asynchronous states and synchronous states. In these cases, 
partial locking is due to the inhomogeneity in intrinsic frequencies. 
For some oscillators, the coupling strength is strong enough to make the oscillators locked to the coupling force by overcoming the desynchronizing effect of frequency difference, but for others it is not strong enough and they drift. 

However, there can be desynchronizing factors other than ones intrinsic to uncoupled oscillators. Recent studies of networks of real systems show connection topology and coupling strength can be far from homogeneous \cite{strogatz2001,albert2002,amaral2000,song2005}. Especially, in many real systems including the world-wide web, the Internet, social and biological networks, the number of connections per node or the total coupling strength per node follows scale-free distribution (or power-law distribution) $P(x) \sim x^{-\gamma}$ \cite{strogatz2001,albert2002}.  Theoretical studies have shown that the inhomogeneity in coupling can induce asynchronous states or make synchronization harder to achieve \cite{golomb2000,denker2004,nishikawa2003,motter2005,zhou2006,twko2007}. 

In this paper, we study the dynamics of oscillators with coupling strength which has a scale-free distribution. 
It is shown that partially locked states can occur due to the cooperation of the inhomogeneity and the coupling function. In contrast to the previously studied partially locked states in systems with distributed intrinsic frequencies, the partially locked states due to coupling inhomogeneity can be bistable with synchronous states.
Using a self-consistency argument, we analytically obtain various values of the partially locked states.

\section{Model and Simulations}
To focus on the coupling inhomogeneity effect, we consider systems of coupled identical limit cycle oscillators with same type of coupling. In the case of weak coupling, the systems can be reduced to the following phase-only ones \cite{kura,strogatz_physica,kura_review,ermentrout2001}.
\begin{eqnarray}
\dot \theta_i = \omega + \frac{1}{N} \sum_{j=1}^{N}K_{ij} H(\theta_j-\theta_i), ~~i = 1,2, ..., N,
\label{eq_model_0}
\end{eqnarray}
where $\theta_i (t)$ is the phase of oscillator $i$ at time $t$, $\omega$ is the natural frequency of the oscillators, and $N$ is the total number of oscillators. $K_{ij}$ is the coupling strength from oscillator $j$ to oscillator $i$ and $K_{ij} \geq 0$. $H(\theta)$ is the coupling function obtained by the phase reduction method for pair-wise interaction \cite{kura,strogatz_physica,kura_review,ermentrout2001}.

Recently, we introduced the following mean-field model as an approximation of the model of Eq. (\ref{eq_model_0}) \cite{twko2007}.
\begin{eqnarray}
\dot \theta_i &=& \omega + \frac{K_i}{N} \sum_{j=1}^{N} H(\theta_j-\theta_i), ~~i = 1,2, ..., N,
\label{eq_model_1}
\end{eqnarray}
where $K_i \;(>0)$ corresponds to the average coupling strength to oscillator $i$.
This model is a simple generalization of the Kuramoto model \cite{kura,strogatz_physica,kura_review} where $K_i$ is the same for all the oscillators, and the coupling inhomogeneity is incorporated in $K_i$. Due to the mean-field coupling, this model, like the Kuramoto model, is easy to simulate and analyze.
In the following sections, we use this model to study the scale-free coupling inhomogeneity and relate the results with those obtained from the simulations with scale-free networks \cite{albert2002}.

Here, we use $H(\theta) = \sin(\theta-\beta) + c_0$ with $c_0 = \sin\beta$ and $\beta \in [0, \pi/2)$ instead of $H(\theta) = \sin\theta$ of the Kuramoto model \cite{kura,strogatz_physica,kura_review}.
In most of the previous studies with coupled phase oscillators on scale-free networks, models with $H(\theta) = \sin\theta$  have been studied \cite{restrepo_papers,mcgraw2005,oh2007}. But coupling functions of the form $H(\theta) = \sin(\theta-\beta) + c_0$ are more realistic approximations of those obtained from coupled limit cycle oscillators \cite{ermentrout2001}, and $c_0$ can affect the dynamics significantly in the systems with inhomogeneity in the number of inputs or in the coupling strength to an oscillators \cite{twko2007,golomb2000}.  
$H(\theta) = \sin(\theta-\beta) + \sin\beta$ is an approximate coupling function for diffusively coupled oscillators such as gap-junction coupled oscillating neurons with which the coupling is zero when the two oscillators are at the same point in the phase space \cite{ermentrout_mbi}. 

With the coupling function $H(\theta) = \sin(\theta-\beta) + \sin\beta$, $\beta \in [0, \pi/2)$, in-phase synchrony is a solution of Eq. (\ref{eq_model_1}) regardless of the coupling strengths, and the state is locally stable since $H'(0) >0$ (Theorem 3.1 of Ref. \cite{ermentrout1992}).

We perform numerical simulations of Eq. (\ref{eq_model_1}) with $H(\theta) = \sin(\theta-\beta) + \sin\beta$ using a fourth order Runge-Kutta method with time step $\Delta t = 0.01$. Unless noted otherwise, we use near uniformly incoherent initial conditions: $\theta_i (0)$ is chosen randomly from $[0, 2\pi)$ for all $i$. Note that near in-phase synchronous initial conditions lead to the locally stable in-phase synchrony. 
Using the rejection method \cite{numericalC},
we randomly select $K_i$ according to a truncated scale-free distribution $g(K)$.
\begin{eqnarray}
g(K) = \left\{\begin{array}{ll}
C K^{-\gamma} &\mbox{for $K\in [K_{\rm min}, K_{\rm max}]$}, \\
0, &\mbox{otherwise},\\
\end{array}
\right.
\label{eq_dist_K}
\end{eqnarray}
where  $K_{{\rm min (max)}} > 0$ and  $C$ is a normalization factor that satisfies the normalization condition
$\int_{0}^\infty g(K) d K = 1$. 
We obtain
$C= (\gamma-1)\: ({K_{\rm{min}}}^{-\gamma+1}-{K_{\rm{max}}}^{-\gamma+1})$.
Truncation in the distribution is introduced to ensure numerical stability but simulations with larger $K_{\rm max}$ show that untruncated scale-free distribution gives only quantitatively different results. 
After the assigning of all the values of $K_i$, we renumber the oscillators according to the ascending order of coupling strength to clearly see the dynamics dependence on $K_i$.

%*************************************************
%Fig. 1
%\psdraft
\begin{figure}
\centering
\epsfig{figure=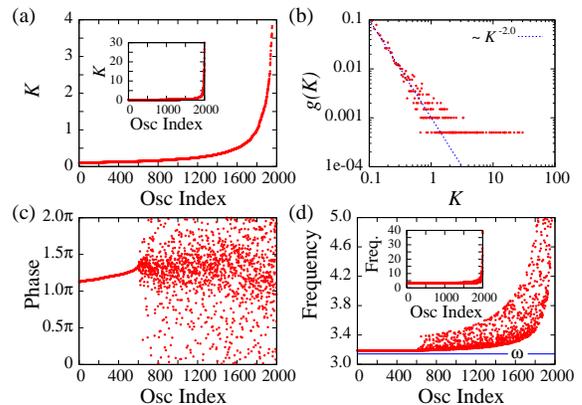, width= 7.5cm}
\caption{Inhomogeneous coupling strength distribution and partial locking for the system of Eq. (\ref{eq_model_1}) with $H(\theta)=\sin(\theta-\beta)+\sin\beta$. $N=2000$ and $\omega=\pi$. (a) Coupling strength $K$ of oscillators. $K_{\rm min} = 0.1$ and $K_{\rm max}=30$ (Eq. (\ref{eq_dist_K})). The inset of (a) shows the entire range of $K$. (b) Coupling strength distribution $g(K)$ for (a). $g(K) \sim K^{-\gamma}$ with $\gamma=2.0$. (c) Phase of oscillators at a certain time after the system reaches a steady state for the case with $\beta = 0.44 \pi$. In this state, oscillators with oscillator index $i$ approximately less than $500$ are locked and others drift. Time averaged order parameter $R \approx 0.657$. (d) Frequency of oscillators for the state of (c). Those oscillators which have the same frequency are locked ones. The insets of (d) shows the entire range of the frequency.  
}
\label{fig_phs}
\end{figure}
%*************************************************

To measure the degree of synchrony, we calculate 
\begin{eqnarray}
 R e^{i \Theta} = \frac{1}{N} \sum_{j=1}^N e^{i \theta_j}.
\label{eq_op}
\end{eqnarray}
$R$ is the order parameter showing the degree of synchrony: $R$ is between $0$ and $1$, with $0$ meaning uniform incoherence and $1$ in-phase synchrony. In the simulations, $R$ approaches a stationary value with small fluctuations and the time average of $R$ is calculated. 

Figure \ref{fig_phs} shows simulation results of Eq. (\ref{eq_model_1}).
In Figs. \ref{fig_phs}(a) and (b), coupling strength $K_i$  and the distribution of $K_i$ are plotted, respectively. $K_{\rm min} = 0.1$ and $K_{\rm max} = 30$ are used in the simulations. 
Figures \ref{fig_phs}(c) and (d) are the snapshot of the phases ($\theta$) and the frequency ($\dot \theta$) of oscillators, respectively. 
While in-phase synchrony is reached from almost all initial conditions for the cases with uniform coupling ($K_i = K$) \cite{watanabe93,watanabe94},  the system evolves to a partially locked state with the coupling inhomogeneity for some range of $\beta$ value (Figs. \ref{fig_phs}(c) and (d)). The oscillators are divided into a phase-locked group giving a continuous line in the figures and drifting group giving scattered dots.
Because the oscillator indices are renumbered according to the order of the coupling strength, the locked oscillators and the drifting ones are clearly distinguished. With the coupling strength distribution treated in this study, it is observed that oscillators with small $K_i$ are locked (Figs. \ref{fig_phs}(a), (c), and (d)).  
 
This is a new phenomenon.
While most of partially locked states discussed previously were due to the inhomogeneity in the natural frequencies of oscillators \cite{kura,strogatz_physica,kura_review}, those of this study are mainly due to the coupling inhomogeneity. Without the inhomogeneity,  the states cannot exist. The formation of partially locked states in this system may be related to the formation of so-called chimera states in nonlocally coupled identical oscillators with a similar coupling function \cite{kuramoto2002,shima2004,abrams2004,abrams2006}. In chimera states which can be classified as partially locked states, phase-locked oscillators spatially coexist with drifting ones. In those systems, there is no coupling inhomogeneity and thus $c_0$ of the coupling function has no significant role. In both of the cases, partially locked states occur when $\beta$ is near $\pi/2$ \cite{kuramoto2002,shima2004,abrams2004,abrams2006}. We can see more of the similarity in the next section through analysis and simulation.
Note also that the partially locked states are bistable with in-phase synchronous states. The bistability between a chimera state and an in-phase synchronous state is a similar one \cite{kuramoto2002,shima2004,abrams2004,abrams2006}.

For $c_0 \neq \sin\beta$, in-phase synchrony is not a solution of the system and unless $c_0$ is large enough to induce uniform incoherence, we can obtain  partially locked states with the truncated scale-free distribution or with other distributions \cite{twko2007}.

\section{Analysis}
In this section, we analyze the partially locked states using a self-consistency argument \cite{kura,strogatz_physica,kura_review}.

Before analyzing partially locked states, we need to look into the local stability of uniformly incoherent states, because uniformly incoherent states are solutions of the model and these states can compete with partially locked states.
In recent paper \cite{twko2007}, using population density analysis, we showed that for the model of Eq. (\ref{eq_model_1}) with $H(\theta) = c_0 + \sin(\theta-\beta)$, the eigenvalues $\lambda = \mu-i\nu - i \omega$ determining the stability of an incoherent state satisfy the equations 

\begin{eqnarray}
2 \cos\beta &=& \int_0^\infty  \frac{\mu K g(K) }{\mu^2 + (K c_0 - \nu)^2} \,dK, \label{eq_lambda_r0}\\
2 \sin\beta &=& \int_0^\infty  \frac{K g(K) (K c_0 - \nu)}{\mu^2 + ( K c_0 - \nu)^2} \,dK,  
\label{eq_lambda_i0}
\end{eqnarray}
where $g(K)$ is the distribution for $K$. 

For the coupling strength distribution of Eq. (\ref{eq_dist_K}), these equations become  

\begin{eqnarray}
\cos\beta &=& \frac{C}{2}\int_{K_{\rm min}}^{K_{\rm max}}  \frac{\mu K^{-\gamma+1}}{\mu^2 + (K c_0 - \nu)^2} \,dK, 
\label{eq_lambda_r1} \\
\sin\beta &=& \frac{C}{2}\int_{K_{\rm min}}^{K_{\rm max}} \frac{K^{-\gamma+1} (K c_0 - \nu)}{\mu^2 + ( K c_0 - \nu)^2} \,dK.  
\label{eq_lambda_i1}
\end{eqnarray}
In the limit of $\mu \rightarrow 0^+$, we can find critical $c_0$ above which uniformly incoherent state is obtained \cite{twko2007}. If we take the conditions $c_0 K_{\rm min} - \nu < 0$ and $c_0 K_{\rm max} - \nu > 0$  without which Eqs. (\ref{eq_lambda_r1}) and  (\ref{eq_lambda_i1}) cannot be satisfied, the equations become the following in the limit.

\begin{eqnarray}
\cos\beta &=&  \;\frac{\pi C}{2 c_0} \left ( \frac{\nu}{c_0}\right)^{-\gamma+1}, 
\label{eq_cos_final}
\\
\sin\beta &=& \lim_{\mu \rightarrow 0^+} \frac{C}{2}\int_{K_{\rm min}}^{K_{\rm max}}  \frac{K^{-\gamma+1} (K c_0 - \nu)}{\mu^2 + ( K c_0 - \nu)^2} \,dK.  
\label{eq_sin_final}
\end{eqnarray}

%*************************************************
%Fig. 2
%\psdraft
\begin{figure}
\centering
\epsfig{figure=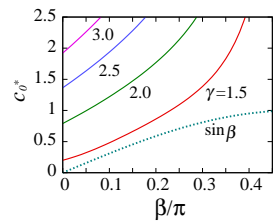, width= 3.78cm}
\caption{
Critical $c_0$ (${c_0}^*$) for the stability of an uniformly incoherent state as a function of $\beta$ for cases with $H(\theta)=c_0+\sin(\theta-\beta)$. Other parameter values are the same as in Fig. 1. The curves are obtained numerically from Eqs. (\ref{eq_cos_final}) and (\ref{eq_sin_final}). When ${c_0} = \sin\beta$, an incoherent state is obtained if  $\sin\beta > {c_0}^*$. For all the values of $\gamma$ of this figure, $\sin\beta < {c_0}^*$ and thus it implies that incoherent states are unstable when ${c_0} = \sin\beta$.
}
\label{fig_c_cr}
\end{figure}
%*************************************************

Figure \ref{fig_c_cr} shows critical $c_0$ (${c_0}^*$) as a function of $\beta$. The critical values denoted by solid lines are obtained numerically from Eqs. (\ref{eq_cos_final}) and (\ref{eq_sin_final}). When $c_0 < {c_0}^*$, incoherent states are unstable. For the case of Eq. (\ref{eq_model_1}) with $c_0 = \sin\beta$ and $\gamma (\geq 1.5$) studied in this paper, $c_0$ is less than ${c_0}^*$ as shown in Fig. \ref{fig_c_cr} and thus the incoherent states are unstable. 
Smaller $\gamma$ gives ranges of $\beta$ values with  $c_0= \sin\beta > {c_0}^*$, but for the simplicity of the discussion we restrict the cases to those with $\gamma \geq 1.5$. Since incoherent states are unstable in this situation, we get states other than incoherent states even when initial conditions are  near-uniformly incoherent states. The system evolves to a partially locked state or an in-phase synchronous state from near-uniformly incoherent states. 

Now, let us analyze the partially locked states.
Kuramoto used self-consistency arguments and derived an equation for the order parameter $R$ to analyze his model in which oscillators with distributed frequencies are globally coupled \cite{kura,strogatz_physica,kura_review}. The idea is to calculate the order parameter by calculating the contribution from locked subpopulation and from drifting subpopulation in stationary states. Because the calculation of the contribution from the subpopulations contains the order parameter, this gives a self-consistent equation for the order parameter. A similar argument with space-dependent order parameter was used by Kuramoto and his colleagues to analyze recently observed chimera states in nonlocally coupled identical oscillators \cite{kuramoto2002,shima2004}. In chimera states, phase-locked oscillators coexist with drifting ones \cite{kuramoto2002,shima2004,abrams2004,abrams2006}. 
We use the same self-consistency argument here to analyze our system. 
In this case $R$ is the same for all the oscillators as in the Kuramoto model, but the coupling term is different for each oscillator as in chimera states because of the dependency of the term on the coupling strength $K_i$. 
This inhomogeneity in the coupling terms causes the splitting of the population into two groups - locked group and drifting group.

We follow the same argument for the analysis of chimera states described in detail in Ref. \cite{abrams2006}.  The main difference is that coupling strength distribution $g(K)$ is used instead of coupling kernel and the order parameter $R$ is the same for all oscillators in our case.  

We assume the limit of infinitely many oscillators and  focus only on stationary states. 
Let $\Omega$ denote the frequency of the population oscillation of Eq. (\ref{eq_op}) after the system approaches a stationary state and $\phi = \theta - \Omega t$ represent the phase of oscillators relative to the average oscillation. Then we can rewrite Eq. (\ref{eq_model_1}) using the order parameter $Re^{i \Theta}$ defined in Eq. (\ref{eq_op}) as follows. 
\begin{eqnarray}
\dot \phi_i &=& \omega -\Omega + K_i \left [R \sin(\Phi-\phi_i-\beta) + \sin\beta \right], \nonumber \\
&&~i = 1,2, ..., N,
\label{eq_sine_1}
\end{eqnarray}
where $\Phi = \Theta - \Omega t$. 
When the system reaches a stationary state, $R$ and $\Phi$ do not depend on time.  The oscillators with $K_i R > |\omega - \Omega + K_i \sin\beta|$ asymptotically approach a stable fixed point ${\phi_i}^*$ obtained from the following equations.

\begin{eqnarray}
\omega - \Omega + K_i\sin\beta = K_i R \sin\left({\phi_i}^* - \Phi + \beta \right)
\label{eq_locked}
\end{eqnarray}
and a stability condition for the fixed point
\begin{eqnarray}
\cos\left({\phi_i}^* - \Phi + \beta \right) > 0. 
\label{eq_locked_stabcond}
\end{eqnarray}
These oscillators are those which are phase-locked at frequency $\Omega$ in the original frame.

The oscillators with  $K_i R < |\omega - \Omega + K_i \sin\beta|$ drift monotonically. This subpopulation can be described by an invariant probability density $\rho(\phi, K)$ in the stationary state. The invariant probability density should satisfy the condition $\rho(\phi, K) v = {\rm constant}$, where $v$ is the instantaneous frequency $\dot \phi$. Therefore, we get
\begin{eqnarray}
\rho(\phi, K) = \frac{\sqrt{{(\omega - \Omega + K \sin\beta)}^2 - K^2 R^2}}{2 \pi | \omega - \Omega + K \sin\beta + K R \sin(\Phi-\phi-\beta)|}, \nonumber \\
\label{eq_density}
\end{eqnarray}
where the normalization constant is chosen such that $\int_{0}^{2\pi}\rho(\phi,K) \, d\phi = 1$.

In the rotating frame of the population oscillation with the frequency $\Omega$,
the order parameter contribution from locked subpopulation is calculated as follows.
\begin{eqnarray}
&&\int_{D_l} dK\; g(K) e^{i \phi^* (K)}  \nonumber \\
&&= e^{-i \beta}  e^{i \Phi} \int_{D_l} dK\; g(K) \nonumber \\
&&\times \frac{\sqrt{K^2 R^2-\left(\Delta+K\sin \beta\right)^2} + i (\Delta + K \sin \beta)}{K R}, \nonumber \\
\label{eq_op_lock}
\end{eqnarray}
where $\Delta \equiv \omega - \Omega$ and  $\mathcal{D}_l$ is the domain with $K R > |\Delta + K \sin\beta|$. 
We use Eqs. (\ref{eq_locked}) and (\ref{eq_locked_stabcond}) to calculate $e^{i \phi^* (K)}$.

The order parameter contribution from drifting subpopulation can be calculated by using population density $\rho(\phi, K)$ of Eq. (\ref{eq_density}).
\begin{eqnarray}
&&\int_{D_d} \int_{0}^{2\pi} d\phi\,dK\; g(K) \rho(\phi, K) e^{i \phi} \nonumber \\
&&= i e^{-i \beta} e^{i \Phi}\int_{D_d} dK\; \frac{g(K)}{KR} \left[ (\Delta + K \sin \beta) \vphantom{\sqrt{\left(\Delta+K\sin \beta\right)^2-K^2 R^2}}\right.  \nonumber \\
&&\left.\vphantom{\int}-{\rm sgn}\left(Z(K,\Delta)\right)\sqrt{\left(\Delta+K\sin \beta\right)^2-K^2 R^2} \;\right] , \nonumber \\
\label{eq_op_drift}
\end{eqnarray}
where $\int_0^{2\pi} d\phi\; \rho(\phi, K) e^{i \phi}$ is calculated using contour integration, ${\rm sgn}(x)$ is the sign function and $Z(K,\Delta) \equiv \Delta + K \sin \beta$. ${\rm sgn}(Z(K,\Delta))$ appears here, because it determines which pole lies inside the contour.  $\mathcal{D}_d$ is the domain with $K R < |\Delta + K \sin\beta|$.

The order parameter $Re^{i\Phi}$ in the rotating frame is the sum of the contributions from locked subpopulation (Eq. (\ref{eq_op_lock})) and from drifting subpopulation (Eq. (\ref{eq_op_drift})). Because $R$ and $\Phi$ are independent of $K$, we obtain

\begin{eqnarray}
R^2 &=& i e^{-i\beta} \left[ \int_{D_{tot}} \frac{g(K)}{K} \left(\Delta + K \sin\beta\right) \: d K  \right. \nonumber \\
&&-i \int_{D_l} \frac{g(K)}{K} \sqrt{K^2 R^2 -\left(\Delta+K\sin\beta\right)^2} \: d K \nonumber \\
&& \left. -\int_{D_d} \frac{g(K)\: {\rm sgn}(Z)}{K} \sqrt{\left(\Delta+K\sin\beta\right)^2-K^2 R^2} \: d K \right], \nonumber \\
\label{eq_R2}
\end{eqnarray}
where $\mathcal{D}_{tot}$ is the total range for $K$.
This gives two independent equations for $R$ and $\Delta$.

$(R=1, \Delta = 0)$ corresponding to the in-phase synchronous state and $(R=0, \Delta)$ corresponding to the uniformly incoherent state are solutions of Eq. (\ref{eq_R2}). But note that this fact does not guarantee the stability of each state.  

We numerically find the solutions $(R < 1, \Delta)$ which correspond to the partially locked states of the system. 
In the cases with coupling strength distribution of Eq. (\ref{eq_dist_K}), numerical simulations of the model show that $K$ with which oscillators are locked are bounded above, $\Delta$ is negative, and ($R-\sin\beta$) is negative. Based on the simulations and the condition for the locking, we can guess that $\mathcal{D}_l = \{K: K_{\rm min} \leq K < K_l \equiv \frac{\Delta}{R-\sin\beta} \}$ and $\mathcal{D}_d = \{K: K_l < K \leq K_{\rm max} \}$. In addition, $Z(K,\Delta)=\Delta+K \sin\beta$ is positive in the domain $\mathcal{D}_d$.  
Using the observation from simulations, we numerically obtain $R$ and $\Delta$ from Eq. (\ref{eq_R2}).

We also compute the fraction of drifting oscillators.
\begin{eqnarray}
f_{drift} &\equiv& \frac{N_{drift}}{N} \nonumber \\
&=& \int_{D_d} g(K) dK \nonumber \\
&=& \int_{K_l}^{K_{\rm max}} g(K) dK \nonumber \\
&=& \frac{{K_l}^{-\gamma+1}-{K_{\rm max}}^{-\gamma+1}}{{K_{\rm min}}^{-\gamma+1}-{K_{\rm max}}^{-\gamma+1}}.
\label{eq_f_drift}
\end{eqnarray}

\begin{figure}
\centering
\epsfig{figure=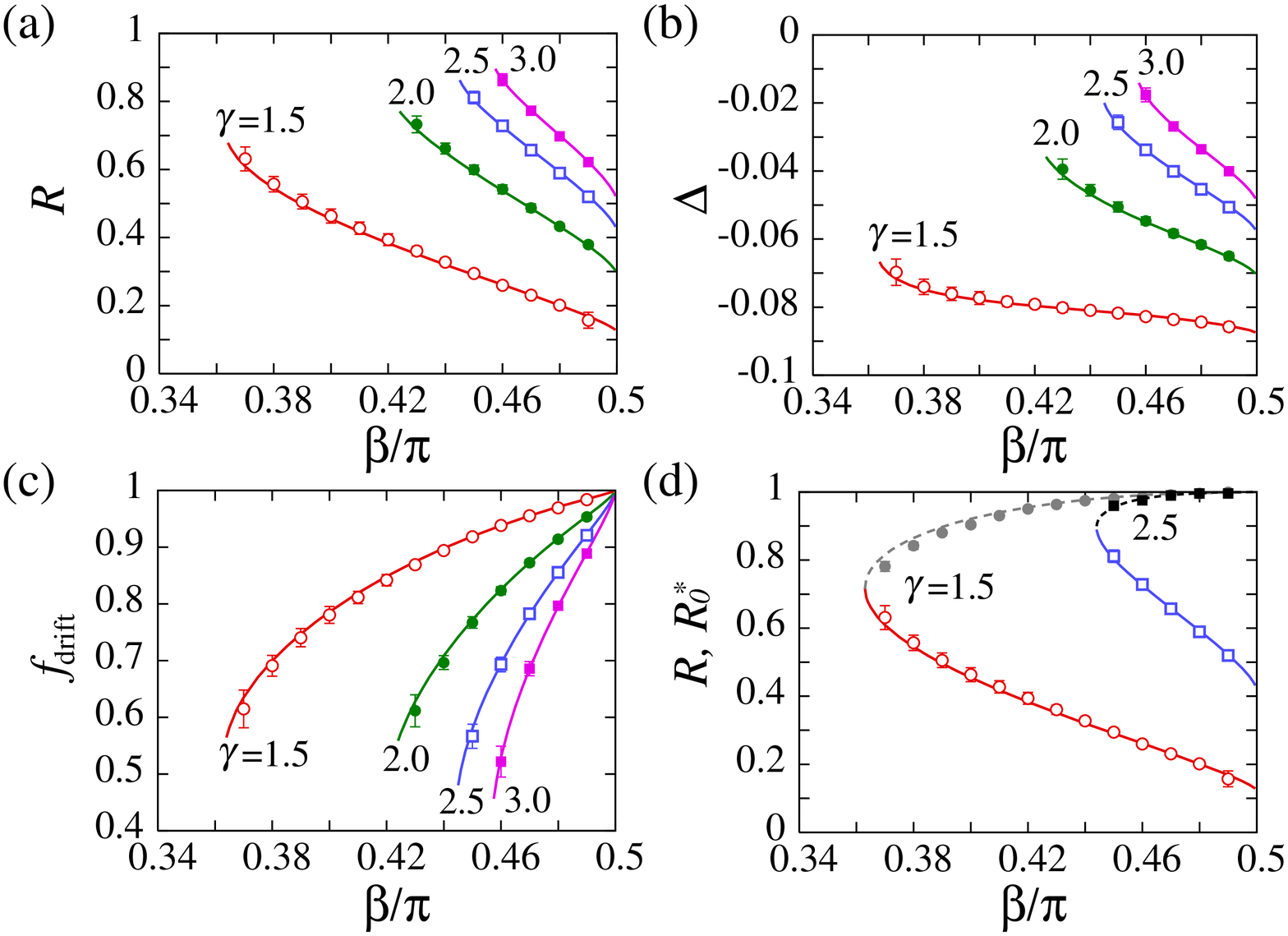, width= 7.5cm}
\caption{
Quantities of partially locked states for different values of  $\gamma$ as a function of $\beta$.
(a) Order parameter $R$.  (b) $\Delta = \omega -\Omega$. (c) Fraction of drifting oscillators.
In (a), (b), and (c), the symbols represent the average values of each quantity which are obtained by first averaging the quantity over time and then over simulations with different $\{K_i\}$ configurations and initial conditions. 
The fluctuations in the time series of each quantity are small compared to the value of each quantity. The curves of (a) and (b) are numerically obtained from Eq. (\ref{eq_R2}).  The curves of (c) are obtained from Eq. (\ref{eq_f_drift}) using $R$ and $\Delta$ from Eq. (\ref{eq_R2}).
The error bars indicate the standard deviation of the time-averaged quantities. 
(d) Initial conditions that lead to partially locked states. 
The open symbols and the solid curves are for $R$ as in (a). The filled symbols denote the value of critical initial order parameter ${R_0}^*$ above which the initial conditions evolve to in-phase synchronous states. Each dashed curve represents the other nontrivial branch of the solution of Eq. (\ref{eq_R2}) for each $\gamma$. See the text for the initial conditions of (d). 
Other parameter values are the same as in Fig. 1.
}
\label{fig_analysis}
\end{figure}

In Figs. \ref{fig_analysis}(a), (b), and (c), we plot the values of $R$, $\Delta$, and $f_{drift}$ obtained both from simulations and analysis for partially locked states. The analysis (solid lines) shows good agreement with simulation results (symbols). As mentioned, the in-phase state is stable for the entire parameter range of this system (not shown in the figures). We use near-incoherent states as initial conditions in simulations and get partially locked states for $\beta \geq \beta^*$ which depends on $\gamma$. For $\beta < \beta^*$, the system reaches to in-phase synchronous state. Simulations and analysis show that $\beta^*$ becomes larger as $\gamma$ increases. This is consistent with the fact that as $\gamma$ increases, the coupling distribution becomes that of the case with uniform coupling strength and in-phase synchronous states are asymptotically reached from almost all initial conditions in the case with uniform coupling strength \cite{watanabe93,watanabe94}. The partially locked states exist for $\beta$ values near $\pi/2$, which is similar parameter range for the existence of chimera states in the systems with nonlocal coupling \cite{abrams2004,abrams2006}. As for the chimera states, the transition points to partially locked states appear abruptly. This contrasts to the continuous transition between partially locked states and asynchronous states in systems with distributed frequencies \cite{kura,strogatz_physica,kura_review}.

Note that this system has bistability between an in-phase synchronous state and an partially locked state. 

To see how the selection of the state between in-phase state and partially locked state depends on the initial conditions in the presence of inhomogeneity \cite{wiley2006}, we look at the asymptotic state as a function of the initial conditions. The initial values for $\theta_i$ are randomly chosen from $[0, 2\pi r)$, where $r \in  [0, 1]$. The corresponding initial value of $R$ is given by $R_0=\frac{1}{2\pi r} \left|\int_{0}^{2\pi r} e^{i\theta}d\theta\right| = \frac{\sin(\pi r)}{\pi r}$.   
We scan the interval $[0,1]$ for $r$ with step size $0.025$ to find $r$ below which the system goes to the in-phase synchronous state with $0.025$ intervals. We denote the $r$ value $r^*$. We average the corresponding $R_0$ values ($R^*$ values) over states from 10 different configurations. In Fig. \ref{fig_analysis}(d),
the filled symbols show the $R^*$ values. In the simulations, we obtain partially locked states for $R \leq R^*$ and in-phase synchronous states for  $R > R^*$. No other states are obtained in our simulations. Open symbols and solid curves are the same ones as in Fig. \ref{fig_analysis}(a). Each dashed curve represents the other nontrivial branch of the solution of Eq. (\ref{eq_R2}). The values of $R^*$ are close to but below the dashed curves. The states for the dashed curves seem to be unstable and seem to act as basin boundaries for at least this type of initial conditions. 
The fact that the $R^*$ values are close to $1$ 
shows that the basin of attraction for the partially locked states is relatively large compared to that of the in-phase synchronous states  and thus the partially locked states can occur rather naturally in this system.
Synchronous states are stable even with the inhomogeneity but the inhomogeneity makes the size of the sync basin smaller \cite{wiley2006}.

%*************************************************
%Fig. 3
%\psdraft
%*************************************************

\section{Simulation Results with Truncated Scale-free Networks}
The results of previous sections are applicable in systems of coupled oscillators on networks.  

\begin{eqnarray}
\dot \theta_i &=& \omega + \frac{K}{N} \sum_{j=1}^{N} A_{ij}\left[\sin(\theta_j-\theta_i-\beta) + \sin\beta\right] \nonumber \\
&&~i = 1,2, ..., N,
\label{eq_sf_model}
\end{eqnarray}
where oscillator $i$ is influenced by $k_i$ neighbors with coupling strength $K$ according to a coupling topology described by an adjacency matrix $A$. $k_i$ is called the degree of $i$. We take the element of adjacency matrix $A_{ij}=1$, if oscillator $j$ influences oscillator $i$, and $A_{ij}=0$ otherwise.

We consider the cases of truncated scale-free networks \cite{albert2002} where the degree distribution of $k_i$, $P(k)$, is given by 
\begin{eqnarray}
P(k) = \left\{\begin{array}{ll}
C k^{-\gamma} &\mbox{for $k\in [k_{\rm min}, k_{\rm max}]$},\\
0, &\mbox{otherwise}.\\
\end{array}
\right.
\label{eq_ki}
\end{eqnarray}
When oscillators are randomly coupled to others and $k_{\rm min}$ is sufficiently large, we can use the following approximation for Eq. (\ref{eq_sf_model}). 
\begin{eqnarray}
\sum_{j=1}^N A_{ij} H(\theta_j-\theta_i) \approx \frac{k_i}{N}\sum_{j=1}^N H(\theta_j-\theta_i). 
\label{eq_sum_approx}
\end{eqnarray}
With this, Eq. (\ref{eq_sf_model}) is approximately equivalent to
\begin{eqnarray}
\dot \theta_i &=& \omega + \frac{K k_i}{N^2} \sum_{j=1}^{N} \left[\sin(\theta_j-\theta_i-\beta) + \sin\beta\right],
\label{eq_sf_approx_model}
\end{eqnarray}
which is Eq. (\ref{eq_model_1}) with $K_i$ being  $\frac{K k_i}{N}$.

%*************************************************
%Fig. 4
%\psdraft
\begin{figure}[h]
\centering
\epsfig{figure=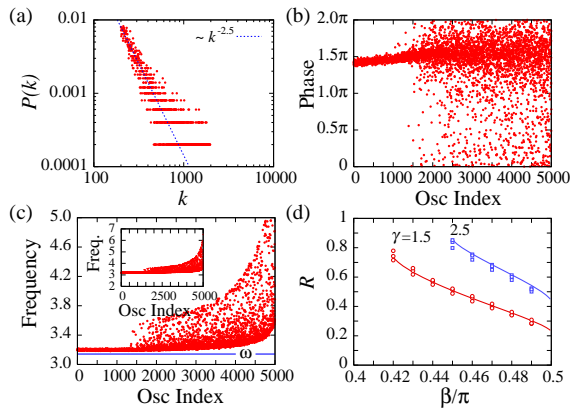, width= 7.5cm}
\caption{
Inhomogeneous degree distribution and partial locking for the system of Eq. (\ref{eq_sf_model}) with $H(\theta)=\sin(\theta-\beta)+\sin\beta$. $\omega = \pi$ and $K=1$.
(a) Degree distribution $P(k) \sim k^{-\gamma}$ with $\gamma=2.5$. 
$N=5000$, $k_{\rm min} = 200$ and $k_{\rm max}=2000$ (Eq. (\ref{eq_dist_K})). 
(b) Phase of oscillators at a certain time after the system reaches a steady state for the case with  $\beta = 0.46 \pi$. (c) Frequency of oscillators for (b).  (d) Order parameter $R$. The symbols represent numerically obtained time-averaged order parameter for each simulation. The curves denote the values obtained from the analysis (Eq. (\ref{eq_R2})). 
}
\label{fig_sf_phs}
\end{figure}

We simulate Eq. (\ref{eq_sf_model}) with networks following a given degree distribution $P(k)$ of Eq. (\ref{eq_ki}). The networks are generated as follows.   
Using the similar method of Section II, we randomly select a positive integer $k \in [k_{\rm min}, k_{\rm max}]$ and assign it to an oscillator among the $N$ oscillators as the degree of the oscillator. We randomly select $k$ oscillators as the neighbors. After the network is generated, the oscillators are renumbered according to the ascending order of the degree. For simplicity of generating networks, we use directed networks but bidirectional networks do not change the results significantly.  

Figure \ref{fig_sf_phs} shows the simulation results of Eq. (\ref{eq_sf_model}).
Figure \ref{fig_sf_phs}(a) shows the degree distribution of the network.
Figures \ref{fig_sf_phs}(b) and (c) are the snapshots of the phases and the frequency of oscillators, respectively.  
As in the cases with inhomogeneous coupling strength shown previous sections, this system also shows similar states. In this system, the states have near locked oscillators (Figs. \ref{fig_sf_phs}(b) and (c)). We compare the simulation results of this system with those obtained from Eq. (\ref{eq_R2}) using the fact $K_i$ corresponds to $\frac{K k_i}{N}$. Figure (\ref{fig_sf_phs})(d) shows good agreement between them.

\section{Summary and Conclusions}
In summary, we have investigated coupled identical oscillators with scale-free distribution of coupling strength and found that partially locked states can occur due to the inhomogeneity and the coupling function. Various quantities of the partially locked states have been computed through a self-consistency argument. This study contrasts with the previous studies in the fact that the coupling inhomogeneity instead of the frequency inhomogeneity is the main cause of partial locking and partially locked states can be bistable with synchronous states.
Our findings may help further understanding of synchronous behavior on inhomogeneous networks. 

\section*{Acknowledgment}
This work was supported by National Science Foundation grant DMS05135.

%==================================================================

\end{document}